# Phonon- and defect-limited electron and hole mobility of diamond and cubic boron nitride: a critical comparison

*Nocona Sanders, Emmanouil Kioupakis**

Department of Materials Science and Engineering, University of Michigan, Ann Arbor MI 48109, United States

ABSTRACT

Diamond and cBN are two of the most promising ultra-wide-band-gap (UWBG) semiconductors for applications in high-power high-frequency electronic devices. Yet despite extensive studies on carrier transport in these materials, there are large discrepancies in their reported carrier mobilities. In this work, we investigate the phonon- and dopant-limited electron and hole mobility of cBN and diamond with atomistic first-principles calculations in order to understand their fundamental upper bounds to carrier transport. Our results show that although the phonon-limited electron mobilities are comparable between cBN and diamond, the hole mobility is significantly lower in cBN due to its heavier hole effective mass. Moreover, although lattice scattering dominates the mobility at low doping, neutral impurity scattering becomes the dominant scattering mechanism at higher dopant concentrations due to the high dopant ionization energies. Our analysis provides critical insights and reveals the intrinsic upper limits to the carrier mobilities of diamond and cBN as a function of doping and temperature for applications in high-power electronic devices.

Diamond and cBN are two promising ultra-wide-band-gap (UWBG) semiconductors for applications in high-power high-frequency electronic devices. Despite their ultra-wide band gaps (5.4 eV in diamond[1] and 6.4 eV in cBN[2]), both show n-type and p-type dopability.[3] These wide band-gap values lead to high breakdown fields (>10 MV/cm in diamond,[4] > 15 MV/cm in cBN[3]), enabling device operation at high voltages. Additionally, they have the highest known thermal conductivities among all materials, allowing for efficient thermal management in electronic systems.[5,6] These favorable properties have stimulated research in the electronic properties of cBN and diamond towards a path to commercialization.

Despite extensive studies on carrier transport in these materials, however, there are large discrepancies in their reported carrier mobilities. Early reports of the room-temperature electron (~2300 $cm^2/Vs$ [7]) and hole (~2100 $cm^2/Vs$ [8]) mobility in natural diamond lead to record high values for the Baliga figure of merit (over 550,000 $10^6$ $V^2$ $\Omega^{-1}$ $cm^{-2}$),[3] confirming its appeal for efficient high-power electronics. However, subsequent measurements in diamond samples of varying quality report a broad range of carrier mobility values. Hole mobilities of 138-2016 $cm^2/Vs$[9] and 3800 $cm^2/Vs$[10] have been measured in CVD-grown diamond films, while measured electron mobilities range from 70 $cm^2/Vs$ in polycrystalline thin films,[11] to 1800 $cm^2/Vs$ in single-crystal type IIa[12], and to 4500 $cm^2/Vs$ for an undoped homoepitaxial CVD diamond film.[10] The origins of many of these discrepancies have largely been identified: grain boundaries in polycrystalline samples significantly lower the mobility compared to single crystals,[13] whereas the highest reported values for photo-excited carriers have not been reproduced in comparable measurements,[14] indicating that photo-excitation is not an appropriate method to describe the carrier mobility under thermodynamic equilibrium in electronic devices.[15] Theoretical work studying diamond hole mobility at thermodynamic equilibrium finds a maximum value of 2000 $cm^2/Vs$[16], similar to the original experimental reports by Reggiani et al.[8] Recent experimental reports on phosphorus-doped diamond films, which do not use photo-excitation, report electron mobilities of 660 $cm^2/Vs$[17] and 1000 $cm^2/Vs$,[15] both carried out with low doping concentrations (~$10^{16}$ $cm^{-3}$). The measured room-temperature electron mobility of cBN (825 $cm^2/Vs$)[18] is similarly high. However, the value of the room-temperature hole mobility of cBN is less clear, with experimental reports spanning a range of two orders of magnitude (from 2 $cm^2/Vs$ [19] to 500 $cm^2/Vs$ [20]). Until recently, the growth of high-quality samples of cBN has proven difficult. Challenges in cBN film deposition and identification of fine grains often lead to samples that are stressed or consisting of mixed phases, containing the hexagonal BN phase and elemental boron.[21] The wide range of sample qualities and preparations have resulted in an inconsistent range of reported hole-mobility values. It is therefore vital to understand the mechanisms that govern carrier transport in diamond and cBN to elucidate the disparity of their reported mobility values and uncover their true potential in power-electronic devices.

| | cBN | | Diamond | |
|---|---|---|---|---|
| | Electron Mobility ($cm^2/Vs$) | Hole Mobility ($cm^2/Vs$) | Electron Mobility ($cm^2/Vs$) | Hole Mobility ($cm^2/Vs$) |
| Theory, phonon-limited (this work) | 1610 | 80.4 | 1790 | 1970 |
| Natural Diamond | - | - | ~2300[7] | ~2100[8] |
| CVD-grown diamond films | - | - | 4500[10] | 138-2016[9]<br>3800[10] |
| Polycrystalline diamond thin films | - | - | 70[11] | - |
| Single-crystal type IIa diamond | - | - | 1800[12] | - |
| P-doped diamond films | - | - | 660[17]<br>1000[15] | - |
| Single-crystal cBN | 825[18] | 2[19] | - | - |
| cBN thin film | - | 500[20] | - | - |

Table 1. A summary of experimental room-temperature carrier mobility values in cBN and diamond.

In this work, we investigate the intrinsic phonon- and dopant-limited electron and hole mobility of cBN and diamond with atomistic first-principles calculations to understand the fundamental upper bounds to carrier transport in these ultra-wide-band-gap semiconductors. We analyze the mode-resolved electron-phonon coupling matrix elements and scattering rates to elucidate the mechanisms that hinder the carrier mobility. Our calculations show that, while the phonon-limited electron mobilities for the two materials are similarly high, the hole mobility of cBN is lower than that of diamond by a factor of ~25 at room temperature, primarily due to the increased density of states for hole scattering due to the heavier hole effective mass in cBN. Our results suggest that the hole mobility of cBN is much lower than some higher previously reported experimental values measured for thin-film samples on conducting substrates. We also show that the high ionization energy of dopants in both materials leads to significant scattering by neutral impurities, which limit the overall carrier mobility at high dopant concentrations. Our analysis reveals the intrinsic limits to the mobility of diamond and cBN in high-power electronic devices.

Our first-principles approach is based on density functional theory, many-body perturbation theory, density functional perturbation theory, and the efficient interpolation of the electron-phonon coupling matrix elements with the maximally localized Wannier function method. Structural relaxation calculations were done in DFT using the local density approximation (Perdew Zunger parameterization) for the exchange-correlation potential[22,23] within Quantum ESPRESSO[24] using a plane-wave basis and norm-conserving pseudopotentials generated with the fhi98PP code[25] for the valence electrons of B, N, and C. The relaxed lattice constants are in good agreement with experiment (3.615 Å for cBN[26] and 3.567 Å for diamond[27]), being underestimated by only 1.3% for cBN and 0.18% for diamond. DFT and GW calculations were performed with an $8 \times 8 \times 8$ Monkhorst−Pack mesh for the sampling of the first Brillouin zone (BZ). Quasiparticle band structures were calculated using the one-shot $G_0W_0$ method[28] within BerkeleyGW,[29] as this method has proven to produce physically-sound and sufficiently accurate band structures. The DFT and GW eigenvalues are converged to within 2 meV using a plane-wave cutoff of 100 Ry for cBN and 130 Ry for diamond. Quasiparticle band-structure calculations were performed using dielectric-matrix screening cutoff energies of 50 Ry for BN and 34 Ry for diamond. The Hybertsen-Louie generalized plasmon-pole model was employed for the frequency dependence of the dielectric function,[30] and the static remainder approach[31] was used to accelerate the convergence of the summation over unoccupied states. Quasiparticle band structures were interpolated using the maximally localized Wannier function method and the Wannier90 code.[32]

Initial projections for the minimization used the valence *s* and *p* orbitals for boron, nitrogen, and carbon *s* and *p* orbitals for diamond. Phonon frequencies were determined using density functional perturbation theory (DFPT)[33] on a $8 \times 8 \times 8$ BZ sampling grid. Electron-phonon coupling matrix elements were evaluated using the maximally localized Wannier function method[34] within the Electron-Phonon-Wannier (EPW) code,[35] using a carrier density of $10^{18}$ cm$^{-3}$, and interpolated to fine electron and phonon BZ sampling meshes up to $88 \times 88 \times 88$ for cBN and $80 \times 80 \times 80$ for diamond. Polar corrections to the electron-phonon matrix elements were applied for cBN.[36] The phonon-limited carrier mobility was evaluated as a function of temperature with the iterative Boltzmann Transport Equation method[37] for carrier states within a 0.3 eV energy range from the band extrema for cBN and within 0.35 eV for diamond. To account for impurity scattering we employ the semi-analytical Brooks-Herring model for ionized impurities[37–40] and the Erginsoy model for neutral impurities,[41] using the first-principles density-of-states effective masses and dielectric constants as inputs.

We first determine the electronic band structures of cBN and diamond (Fig. 1). Both materials have indirect band gaps, with the valence band maximum (VBM) at Γ, and the conduction band minimum (CBM) at X for cBN, and along the Γ-X high-symmetry path in diamond. We find a band gap of 6.80 eV for cBN and 5.66 eV for diamond, in reasonable agreement with experiment (6.4 eV for cBN,[2] 5.47 eV for diamond[1]). The differences are primarily due to the renormalization of the band gap by the electron-phonon interaction,[42] which is not considered in this work. Whereas the top (bottom) cBN valence (conduction) states have predominantly nitrogen (boron) character, the diamond valence and conduction states are equally composed of the sp3 orbitals of the two carbon atoms. Although the band structures are qualitatively similar, the weaker second-nearest-neighbor overlap of N orbitals leads to heavier hole effective masses in cBN (Table 2), which strongly impact the resulting hole-transport properties. A larger effective mass both reduces the mobility directly, being inversely proportional to the mobility, and indirectly, as it yields a larger density of states and thus a shorter carrier scattering time.

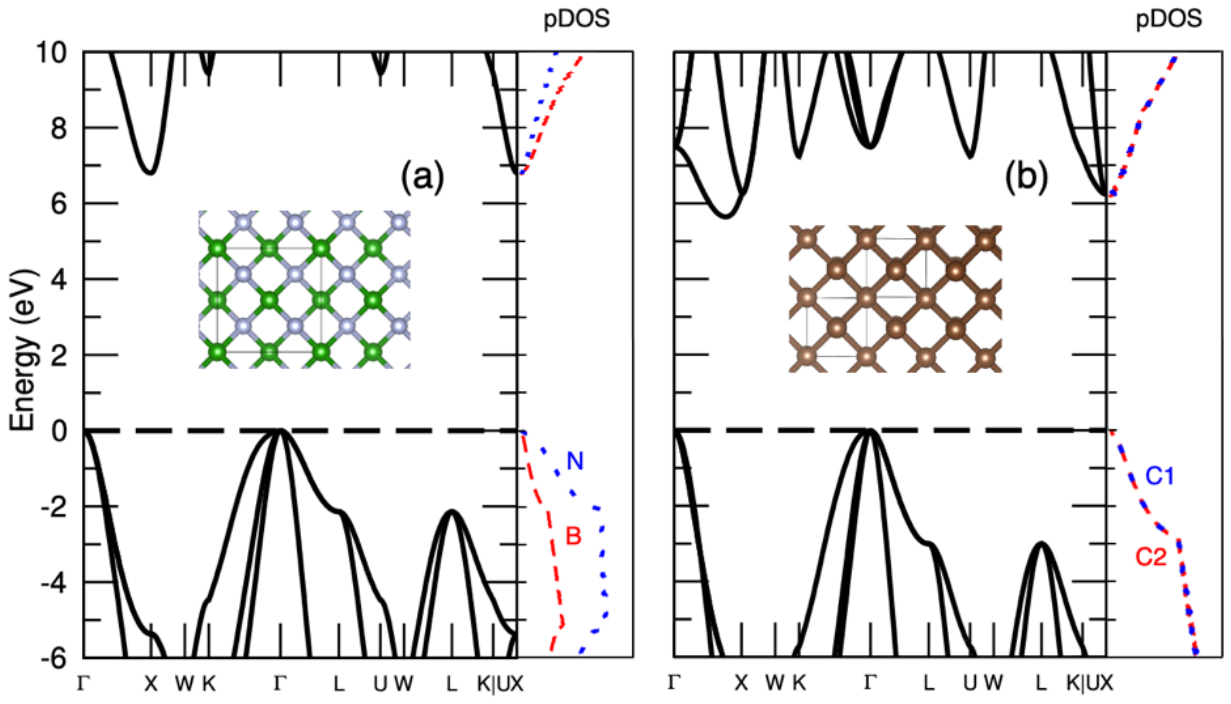

Figure 1. The band structure and atom-projected density of states of (a) cBN and (b) diamond, shown with their crystal structures. While the band gaps are similar, the hole effective masses are heavier in cBN. This is because the cBN valence bands consist primarily of second-nearest-neighbor N orbitals, while the diamond valence band is formed by nearest-neighbor C orbitals.

| **Effective Mass** | **cBN** | | **Diamond** | |
|---|---|---|---|---|
| | This work | Previous theory[43] | This work | Experiment |
| $m^*_{e,l}$ | 1.15 (X → Γ) | 1.20 (X → Γ) | 1.55 (CBM → Γ / X) | 1.4[7] |
| $m^*_{e,t}$ | 0.27 (X → W) | 0.26 (X → W) | 0.27 (X → W) | 0.36[7] |
| $m^*_{hh}$ | 3.00 (Γ → K)<br>1.20 (Γ → L)<br>0.54 (Γ → X) | 3.16 (Γ → K)<br>1.20 (Γ → L)<br>0.55 (Γ → X) | 2.52(Γ → K)<br>0.71 (Γ → L)<br>0.45 (Γ → X) | 1.08[8] |
| $m^*_{lh}$ | 0.50 (Γ → K)<br>1.20 (Γ → L)<br>0.45 (Γ → X) | 0.64 (Γ → K)<br>1.20 (Γ → L)<br>0.54 (Γ → X) | 0.48 (Γ → K)<br>0.71 (Γ → L)<br>0.29 (Γ → X) | 0.36[8] |
| $m^*_{so}$ | 0.27 (Γ → K)<br>0.23 (Γ → L)<br>0.45(Γ → X) | 0.44 (Γ → K)<br>0.36 (Γ → L)<br>0.54 (Γ → X) | 0.33 (Γ → K)<br>0.16 (Γ → L)<br>0.29 (Γ → X) | 0.15[8] |

Table 2. Carrier effective masses of cBN and diamond after GW corrections. Although the electron masses are similar, holes are heavier in cBN.

The phonon dispersions of cBN and diamond are presented in Figure 2. Calculated sound velocities and optical-mode frequencies at Γ are in good agreement with experiment (Table 3). Aside from the expected LO-TO splitting present at Γ in cBN due to the polar crystal structure, the phonon dispersions are qualitatively similar, with comparable phonon frequencies throughout the BZ.

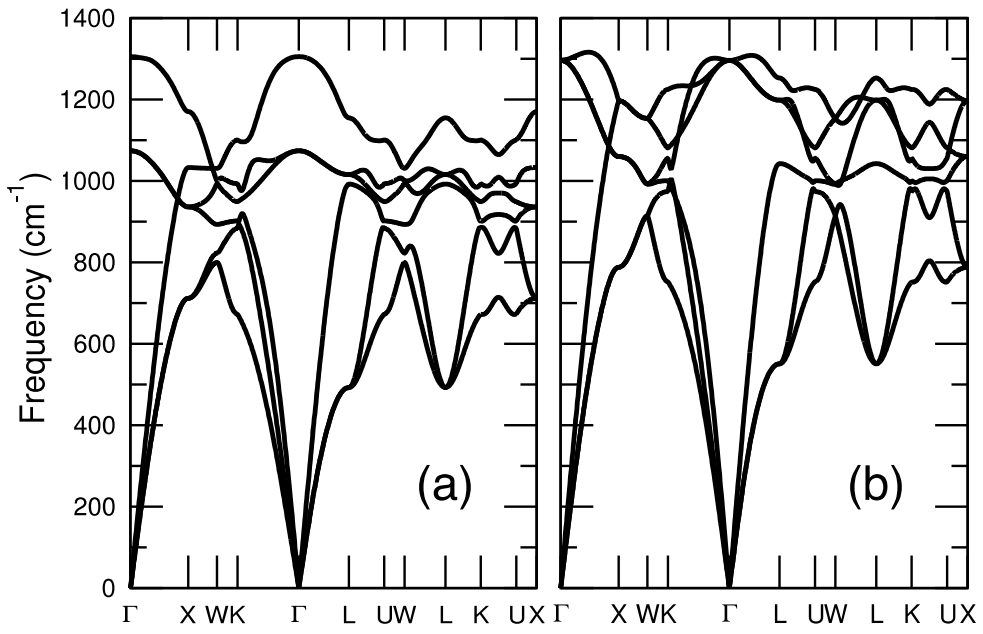

Figure 2. Phonon dispersions of (a) cBN and (b) diamond. Apart from the LO-TO splitting at Γ in polar c-BN, both structures exhibit similar maximum phonon frequencies, sound velocities, and overall band characteristics.

| Property | cBN | | Diamond | |
|---|---|---|---|---|
| | **Theory (this work)** | **Experiment** | **Theory (this work)** | **Experiment** |
| $\omega_{TO}$ (cm$^{-1}$) | 1074 | 1054[a] | 1296 | 1332[b] |
| $\omega_{LO}$ (cm$^{-1}$) | 1305 | 1305[a] | 1296 | 1332[b] |
| $V_{L,(111)}$ (m/s) | 16000 | 14600[c], 14180[c] | 17100 | 19039[d] |
| $V_{T,(111)}$ (m/s) | 9420 | 9570[c], 9072[c] | 11100 | 12300[e] |

[a]Reference [44]
[b]Reference [45]
[c] Polycrystalline - Reference [46]
[d] Reference [46]
[e] Reference [47]

Table 3. Our calculated values for the optical phonon frequencies at Γ and acoustic sound velocities in cBN and diamond compared to experimental measurements.

We next analyze the electron-phonon coupling matrix elements for the topmost valence and lowest conduction band in both materials. We evaluated matrix elements for electrons at the CBM and for holes at the VBM, and for phonon wave vectors along high-symmetry directions of the first BZ (Figs. 3 and 4). In each case, the BZ paths include states occupied by carriers at room temperature. The polar optical (LO) mode couples most strongly both to electrons and holes in cBN, while in diamond the electrons couple most strongly to the LA mode and the holes couple most strongly to the TO modes. However, although carriers in cBN are more strongly coupled to the LO mode, the emission rate of LO phonons is limited due to the high carrier energy required to emit these high-frequency modes, while the absorption rate is limited by the low optical-phonon Bose-Einstein occupation numbers. Similar arguments apply to the scattering of carriers by the optical modes in diamond. Thus, optical-phonon scattering is suppressed, and acoustic modes are expected to be the dominant carrier-scattering phonon modes in both materials, leading to high mobility values.

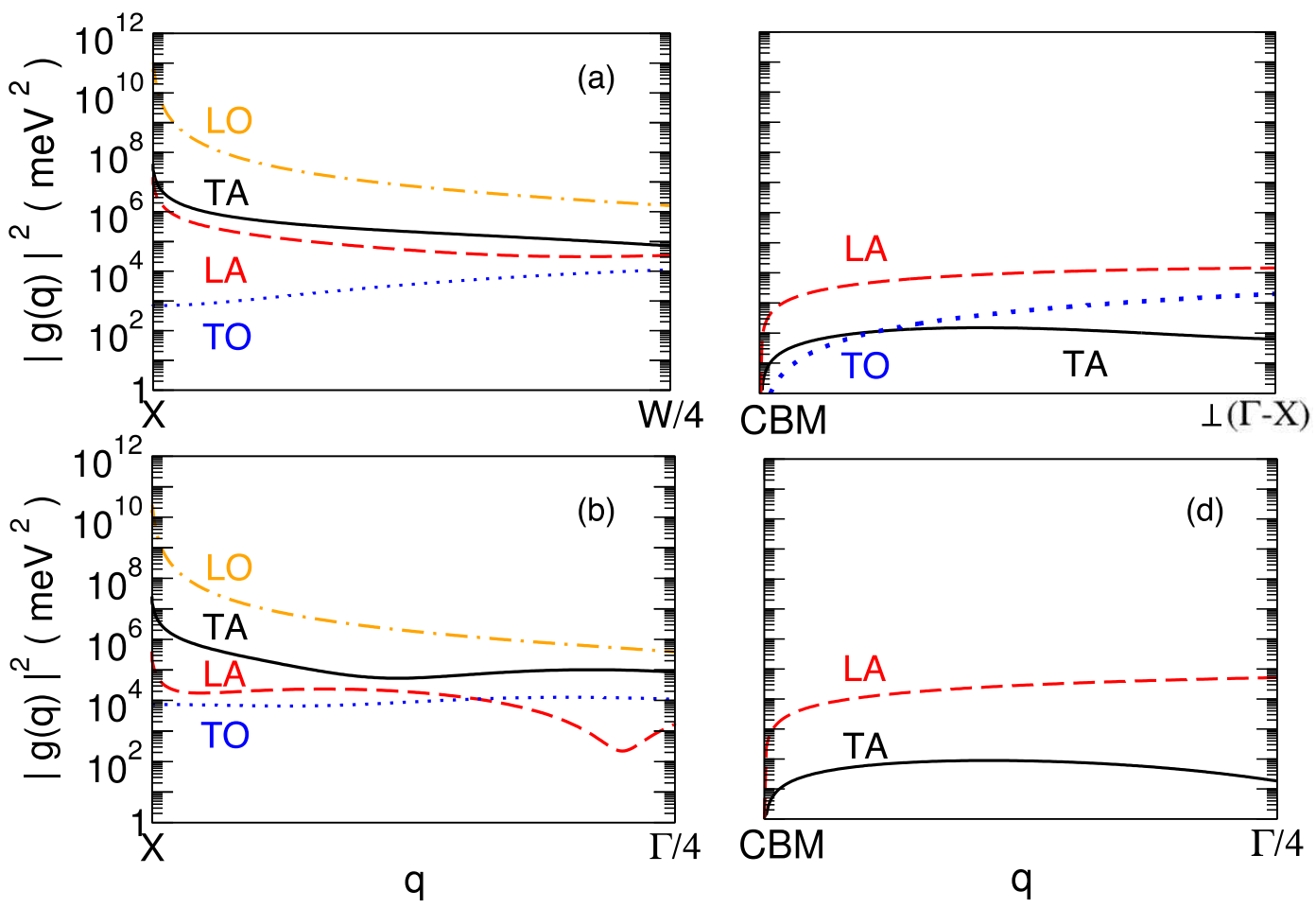

Figure 3. Squared electron-phonon coupling intraband matrix elements for the conduction band of (a)-(b) cBN and (c)-(d) diamond as a function of the phonon wave vector along two perpendicular crystallographic directions. The 1/4 fractions denote a quarter of the distance along the reciprocal-space path to the designated high-symmetry point. The modes that couple most strongly to electrons are the polar optical (LO) modes in c-BN and the LA modes in diamond.

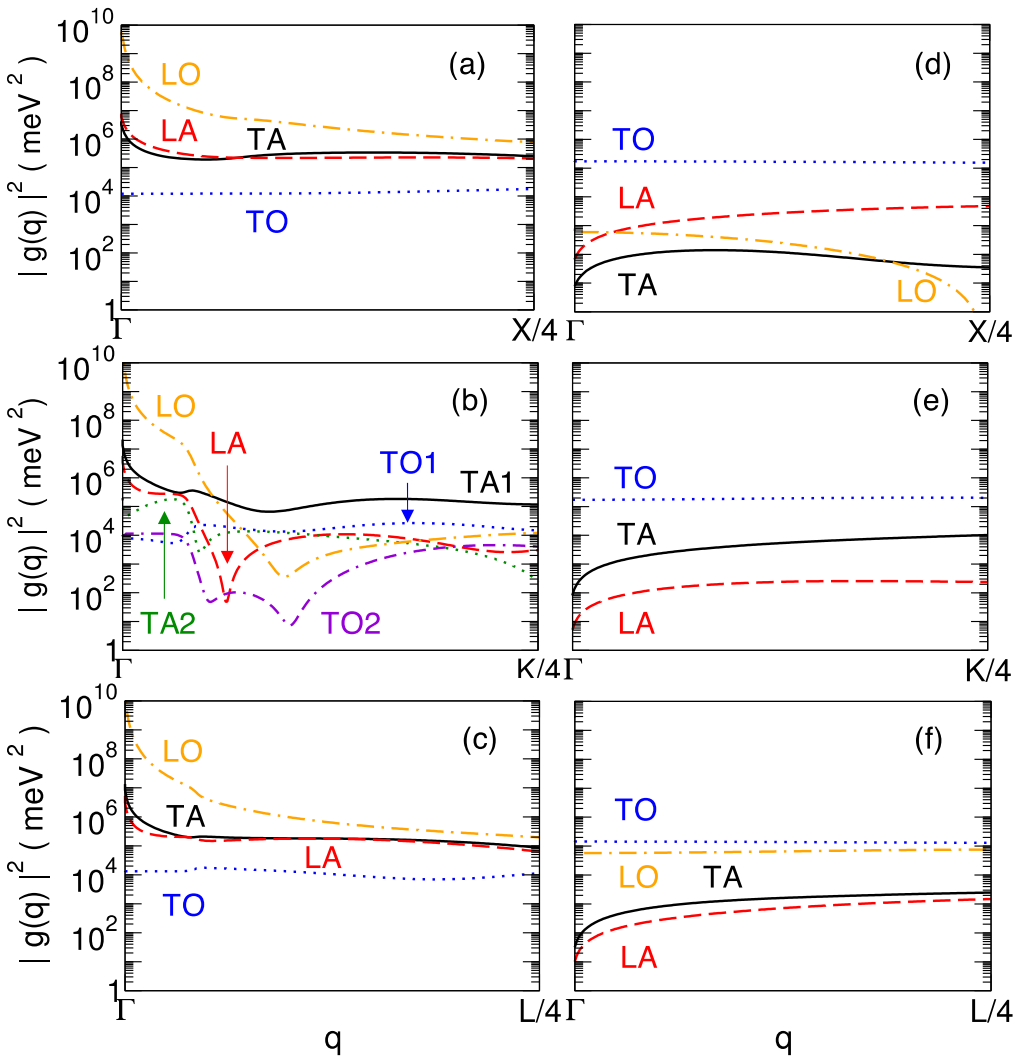

Figure 4. Squared electron-phonon coupling matrix elements for the topmost valence band of (a)-(c) cBN and (d)-(f) diamond. The ¼ fractions denote a quarter of the distance along the reciprocal-space path to the designated high-symmetry point. The modes that couple most strongly to holes for small wave vectors $q$ are the polar optical (LO) modes in c-BN and the TO modes in diamond.

The suppression of optical-phonon electron scattering is confirmed by our calculated phonon-mode-resolved imaginary self-energies (i.e., carrier scattering rates, Fig. 5), which also show that while electron scattering rates are comparable in the two materials, holes scatter at a rate that is 1-2 orders of magnitude faster in cBN. As expected from the discussion in the previous paragraph, the scattering of thermalized carriers (i.e., within a few $k_B T$ from the band extrema) is primarily caused by acoustic modes. On the other hand, optical phonons only become the dominant carrier-scattering modes for carrier energies at least 150 meV away from the band extrema, which have enough thermal energy to emit these high-energy phonons. The similarity of the imaginary self-energies of thermal electrons for the two materials suggests that electron mobilities are similar in the two materials. However, the higher scattering rate of thermalized holes in cBN by 1-2 orders of magnitude caused by the heavy hole effective mass implies that the hole mobility in cBN should be substantially lower than in diamond.

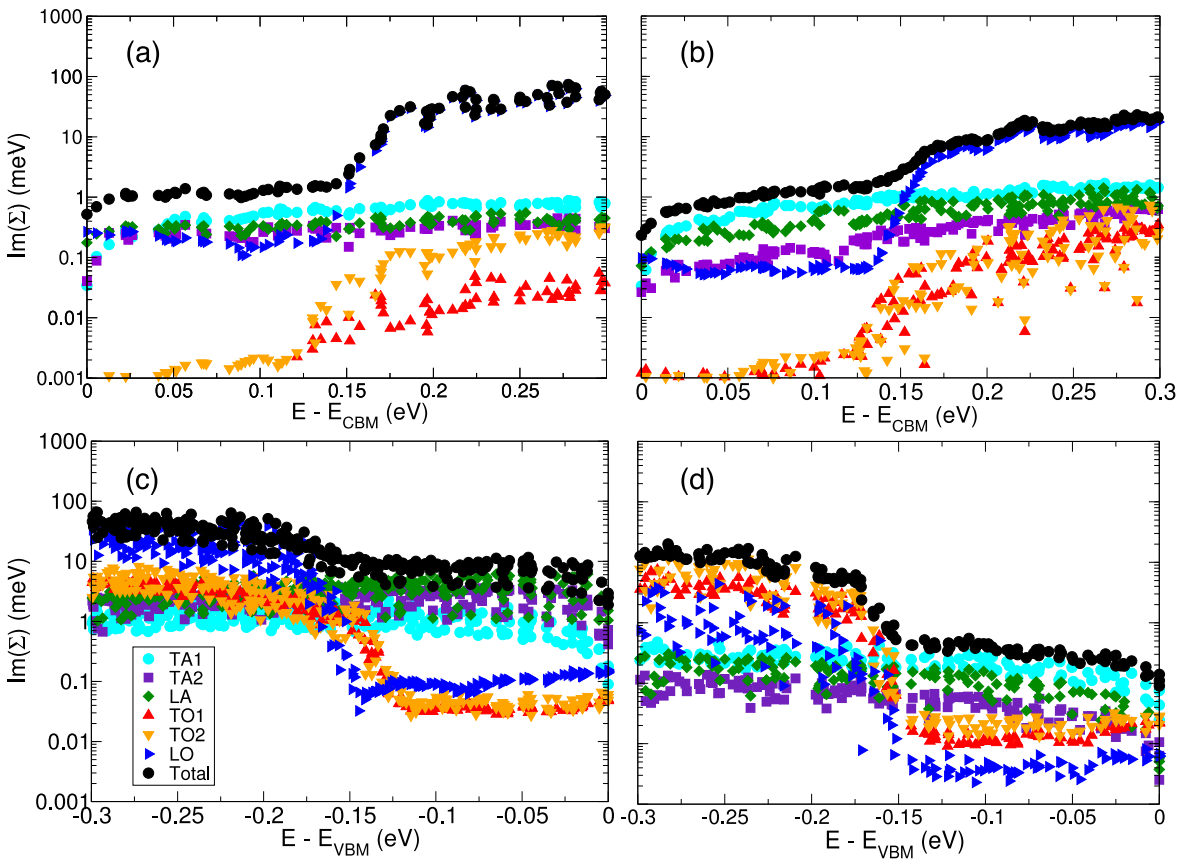

Figure 5. Phonon-mode-resolved imaginary self-energies (i.e., carrier scattering rates) as a function of the carrier energy for (a) electrons in cBN, (b) electrons in diamond, (c) holes in cBN, and (d) holes in diamond. The scattering rates of thermal electrons (i.e., within a few $k_BT$ from the band extremum) are comparable for the two materials, but thermal holes scatter at an 1-2 orders-of-magnitude faster rate in cBN compared to diamond.

The calculated phonon-limited carrier mobilities in cBN and diamond as a function of temperature are shown in Figure 6. Electron and hole mobility values with respect to electron- and phonon-grid densities are converged to within 10 $cm^2/Vs$ and 3 $cm^2/Vs$, respectively. The electron mobility of cBN at room temperature (1610 $cm^2/Vs$) is comparable to that of diamond (1790 $cm^2/Vs$). However, our calculated hole mobility for cBN (80.4 $cm^2/Vs$) is significantly lower than for diamond (1970 cm$^2$/Vs). This difference is attributable to the heavier mass of holes in cBN and the aforementioned increased hole scattering rate, and overall suggests that p-type cBN faces limitations to its performance in high-power electronic devices.

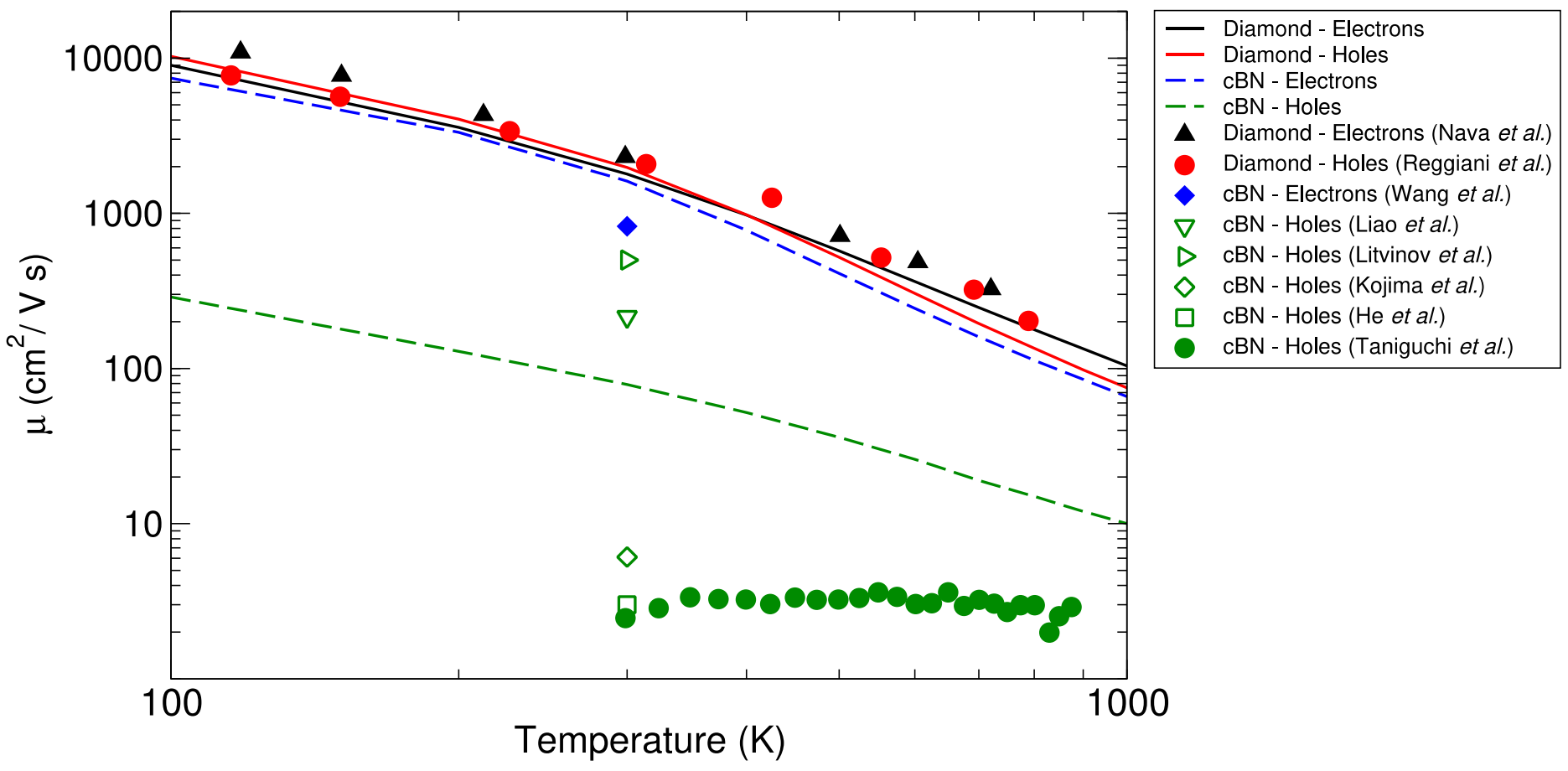

Figure 6. Phonon-limited carrier drift mobilities in cBN and diamond as a function of temperature. The experimental diamond drift mobilities are from natural diamond samples (Nava *et al.*[7] Reggiani *et al.*[8]). Experimental cBN Hall mobilities are from single crystal (Wang *et al.*[18] Taniguchi *et al.*[19]) and thin films. (Litvinov *et al.*,[20] Liao *et al.*,[48] He *et al.*,[49] Kojima *et al.*[50]) Open symbols denote thin-film samples. Although the electron mobility is similar in the two materials, the hole mobility in c-BN is much lower than in diamond. The heavier hole effective mass in c-BN and the resulting larger density of states increase the hole-scattering rates and reduce the hole mobility compared to diamond.

Our mobility results are in overall good agreement with previous experimental measurements (Fig. 6). The experimental data for the diamond carrier mobilities agree well with our calculated values. Our calculated room-temperature phonon-limited electron mobility of cBN (1610 cm$^2$/Vs) is higher than experiment (825 cm$^2$/Vs)[18], possibly due to the presence of imperfections in the experimental sample. However, our calculated value (80.4 cm$^2$/Vs) for the cBN hole mobility is noticeably lower than the highest reported experimental values (500 $cm^2/Vs$).[20] We attribute this discrepancy to the possible conductivity of the silicon substrate used to deposit the characterized cBN thin films in References [20,48], and potentially to the non-optimal material quality in the measured samples that may yield samples with mixed phases. It is possible that boron may incorporate into the silicon substrate used to deposit the cBN thin films, making the resulting silicon sample p-type regardless of any original dopant character. This effect could explain why Litvinov *et al.*[20] measure the same carrier mobility for both n-type and p-type cBN thin films on silicon, especially given that their measured acceptor ionization energy (60 meV) does not match the ionization energy of Be acceptors in cBN ($0.23 \pm 0.03$ eV).[51] Furthermore, the mobility of p-type silicon (480 $cm^2/Vs$)[52] is similar to the reported hole mobility values in cBN by Liao *et al.*[48] and by Litvinov *et al.,*[20] raising the question whether the measured conductivities originate from the substrate rather than the sample. More recent results, which report significantly lower hole-mobility values, were obtained from thin films deposited on insulating quartz[50] or diamond-coated silicon substrates,[49] or single-crystal samples,[19] which eliminate the possibility of the substrate contributing to the measured carrier mobility. Overall, our hole-mobility data are consistent with the lower reported values for hole mobility in cBN single crystals or thin films on insulating substrates.

We also examined how carrier scattering by neutral and charged dopants affects the carrier mobilities of diamond and cBN. We assume that carrier scattering is only caused by the ionized dopants that introduce the free carriers and by unavoidable non-ionized dopants due to the high ionization energies in these two materials. For ionized-impurity scattering we apply the semi-analytical Brooks-Herring model,[37–40]

$$\mu_i = \frac{2^{\frac{7}{2}}\epsilon_s^2\epsilon_o^2(k_BT)^{\frac{3}{2}}}{\pi^{\frac{3}{2}}e^3\sqrt{m_d^*}n_iG(b)},$$

where

$$G(b) = \ln(b+1) - \frac{b}{b+1}$$

and

$$b = \frac{24\pi m_d^*\epsilon_S(k_BT)^2}{e^2h^2n_i},$$

where $\epsilon_S$ is the dielectric constant, $\epsilon_0$ is the vacuum permittivity, $m_d^*$ is the density-of-states mass, and $n_i$ is the density of ionized impurities. To account for neutral impurity scattering, we used the Erginsoy model,[41,53]

$$\mu_n = \frac{\pi^2 m_d^* q^3}{10\epsilon_S\epsilon_0 n_N h^3},$$

where $n_N$ is the density of neutral impurities. Our calculated density-of-states effective mass is 0.48 $m_e$ for electrons in diamond, 1.11 $m_e$ for holes in diamond, 0.69 $m_e$ for electrons in cBN, and 1.19 $m_e$ for holes in cBN. To determine the density of ionized dopants, $n_i$, we use

$$n_i = \sqrt{n_d\frac{1}{4}\left(\frac{2m_d^*k_BT}{\pi\hbar^2}\right)^{\frac{3}{2}}}\, e^{-\frac{E_d}{2k_BT}}$$

where $n_d$ is the dopant density and $E_d$ is the dopant ionization energy. We used experimental values of donor/acceptor energies from the literature of 0.37 eV for boron acceptors in diamond,[54] 0.84 eV for phosphorus donors in diamond,[55] 0.24 eV for silicon donors in cBN,[51] and 0.19 eV for beryllium acceptors in cBN.[51] Our calculated mobility results including dopant scattering are shown in Fig. 7 and compared to experimental values. At low donor/acceptor densities, lattice scattering dominates while at higher doping levels neutral impurity scattering is the major limitation to the mobility. Because of the deep acceptor/donor energies, only a small fraction of the dopants is ionized at room temperature (< 10% in cBN, < 1% in diamond) and thus ionized-impurity scattering plays little role in limiting the overall mobility. Our calculated total mobility agrees well with experimental data for n-type cBN and p-type diamond. Our results for n-type diamond are higher than experiment, presumably because experimental samples contain additional unintentional defects that compensate the phosphorous donors.[56] Moreover, the significantly lower experimental values seen in p-type cBN may be due to material quality or because one of the samples is polycrystalline. Our estimated carrier mean-free paths based on the phonon-limited carrier mobilities and effective masses, ranging from 5 nm for holes in cBN to an order of 100 nm for electrons in cBN and electrons and holes in diamond, are shorter than the typical grain sizes in polycrystalline samples (~1 μm), indicating that grain-boundary scattering is not a limiting factor in reducing the carrier mobility. However, the presence of grain boundaries may introduce additional trapping centers whose concentration increases with decreasing grain size (higher grain surface-to-volume ratio). Such increasing carrier trapping at grain boundaries has been applied to explain the experimentally reported decreasing mobility with decreasing grain size in diamond.[13] Overall we find that defect scattering has a more pronounced effect in the hole mobility of diamond than in cBN, primarily because of the deeper acceptor ionization energies and thus the higher density of non-ionized acceptors that dominate scattering at high doping concentrations.

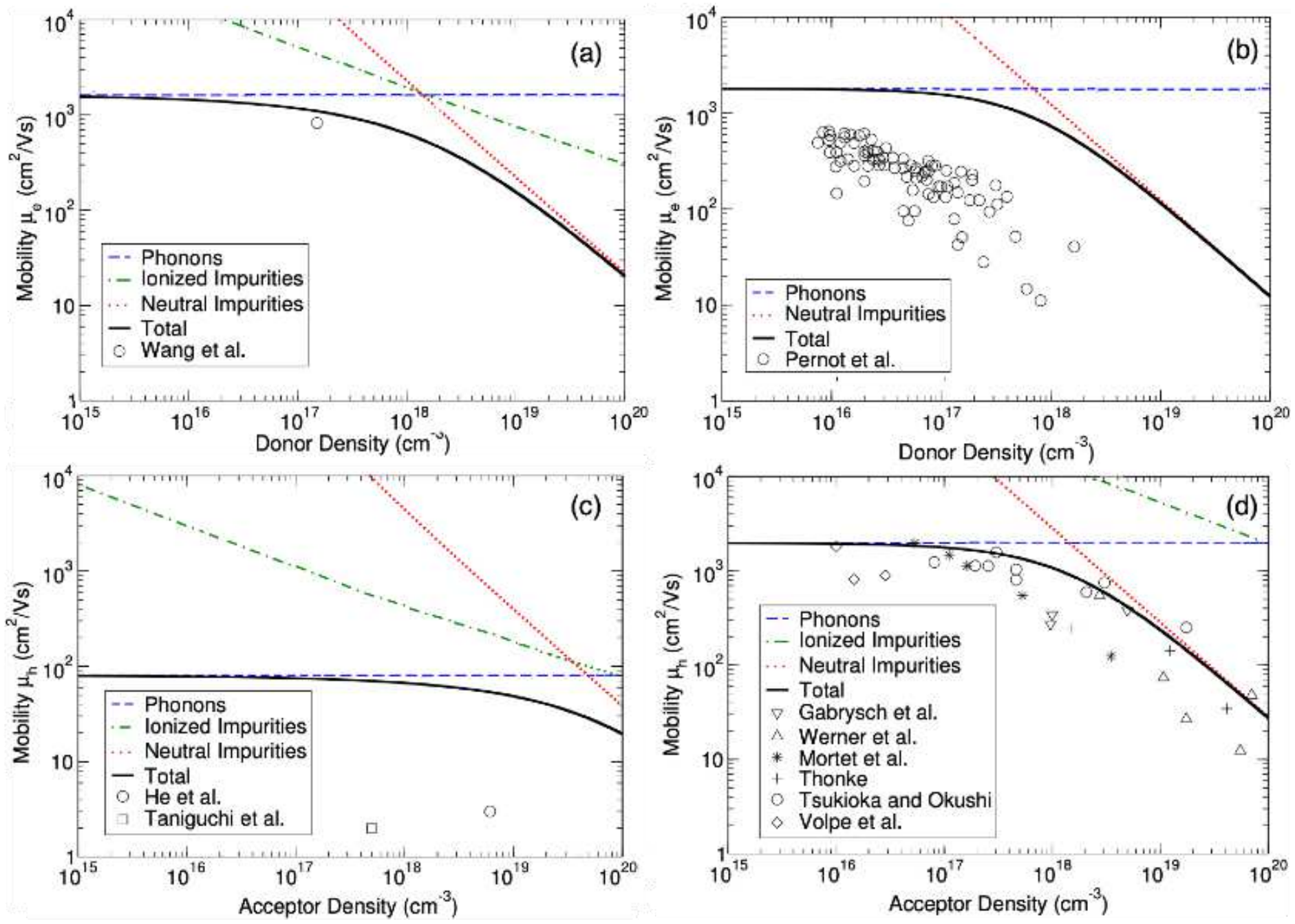

Figure 7. Carrier mobility at 300K as a function of dopant density in (a) n-type, Si-doped cBN, (b) n-type, P-doped diamond, (c) p-type, Be-doped cBN, and (d) p-type, B-doped diamond. Dashed, dotted, and dash-dotted lines show the theoretical contributions from phonon scattering, neutral impurities, and ionized impurities, respectively. Experimental data points are from experimental work by Wang et al.,[18] He et al.,[49] Taniguchi et al.,[19] Pernot et al.,[57] Gabrysch et al.,[58] Werner et al.,[59] Mortet et al.,[16] Thonke,[60] Tsukioka and Okushi,[61] and Volpe et al.[62]

In conclusion, we applied predictive calculations to investigate the intrinsic phonon- and dopant-limited carrier mobilities of cBN and diamond. We found that although the phonon-limited electron mobility of cBN at room temperature is comparable to diamond, the hole mobility for cBN is significantly lower due to its heavier hole effective mass that increases the hole scattering rates. At high doping concentrations scattering is dominated by neutral defect scattering, due to high dopant ionization energies and the resulting high concentrations of non-ionized dopants. Our results explain the physical origin of the lower hole mobility values in cBN compared to diamond, corroborating the findings of recent experiments. Overall, our work uncovers the physical origins and limitations to the carrier mobility of cBN and diamond in high-power electronic devices.

Note added in revision: a recent related computational study reported the phonon-limited mobility values for a series of semiconductors, including diamond and cBN.[63] Although there are differences in the numerical values, the relative magnitudes of the electron and hole mobilities of diamond and cBN are similar to the present work.

AUTHOR INFORMATION

Corresponding Author

*Email: kioup@umich.edu

Funding Sources

This work was supported as part of the Computational Materials Sciences Program funded by the U.S. Department of Energy, Office of Science, Basic Energy Sciences, under Award No. DE-SC0020129. Computational resources were provided by the DOE NERSC facility under Contract No. DE-AC02-05CH11231. Graphics were generated with VESTA.[64]

DATA AVAILABILITY

The data that support the findings of this study are available from the corresponding author upon reasonable request.

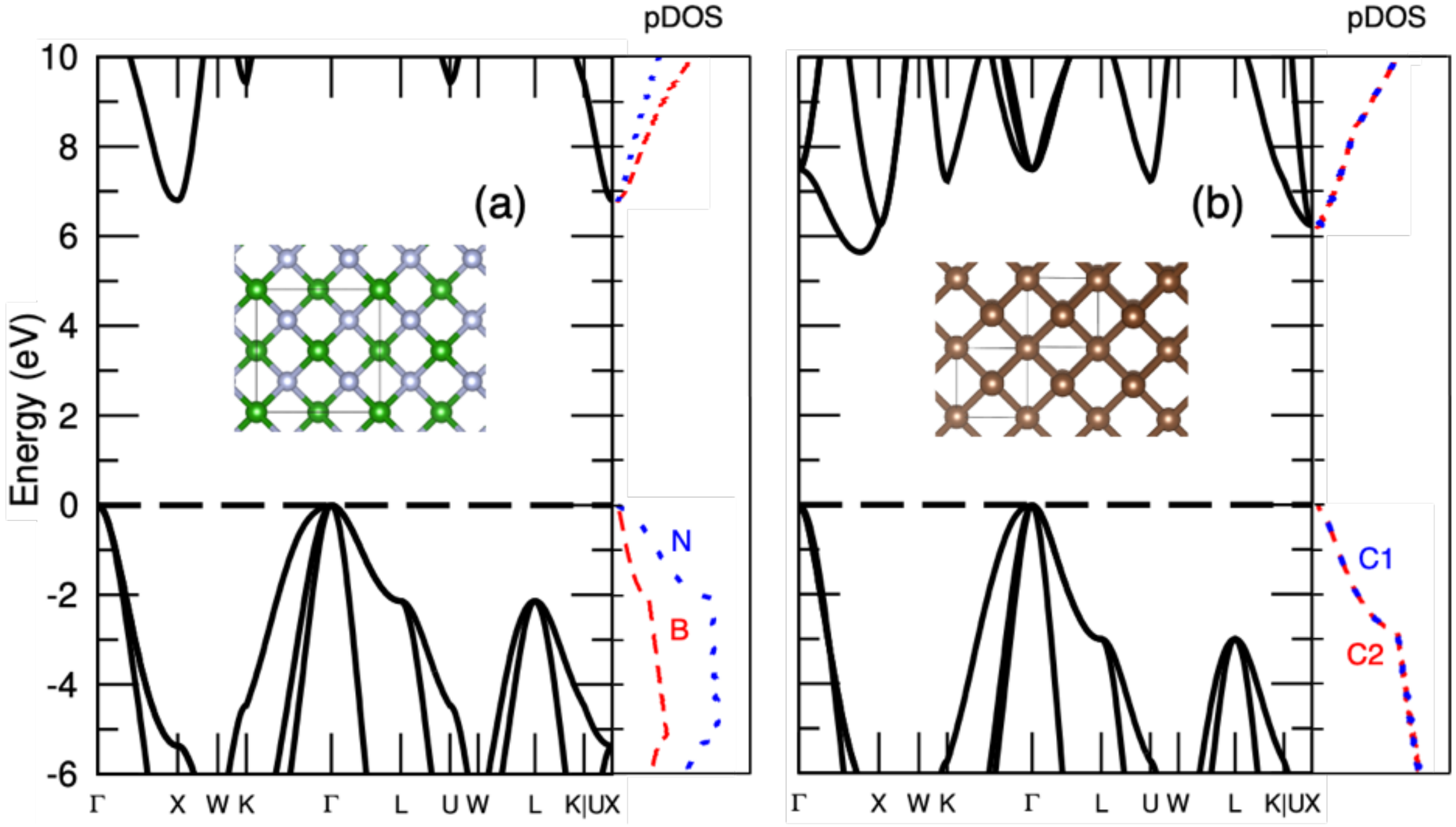
pDOS
pDOS
(a)
(b)
Energy (eV)
10
8
6
4
2
0
-2
-4
-6
N
B
C1
C2
Γ X W K Γ L U W L K|UX
Γ X W K Γ L U W L K|UX

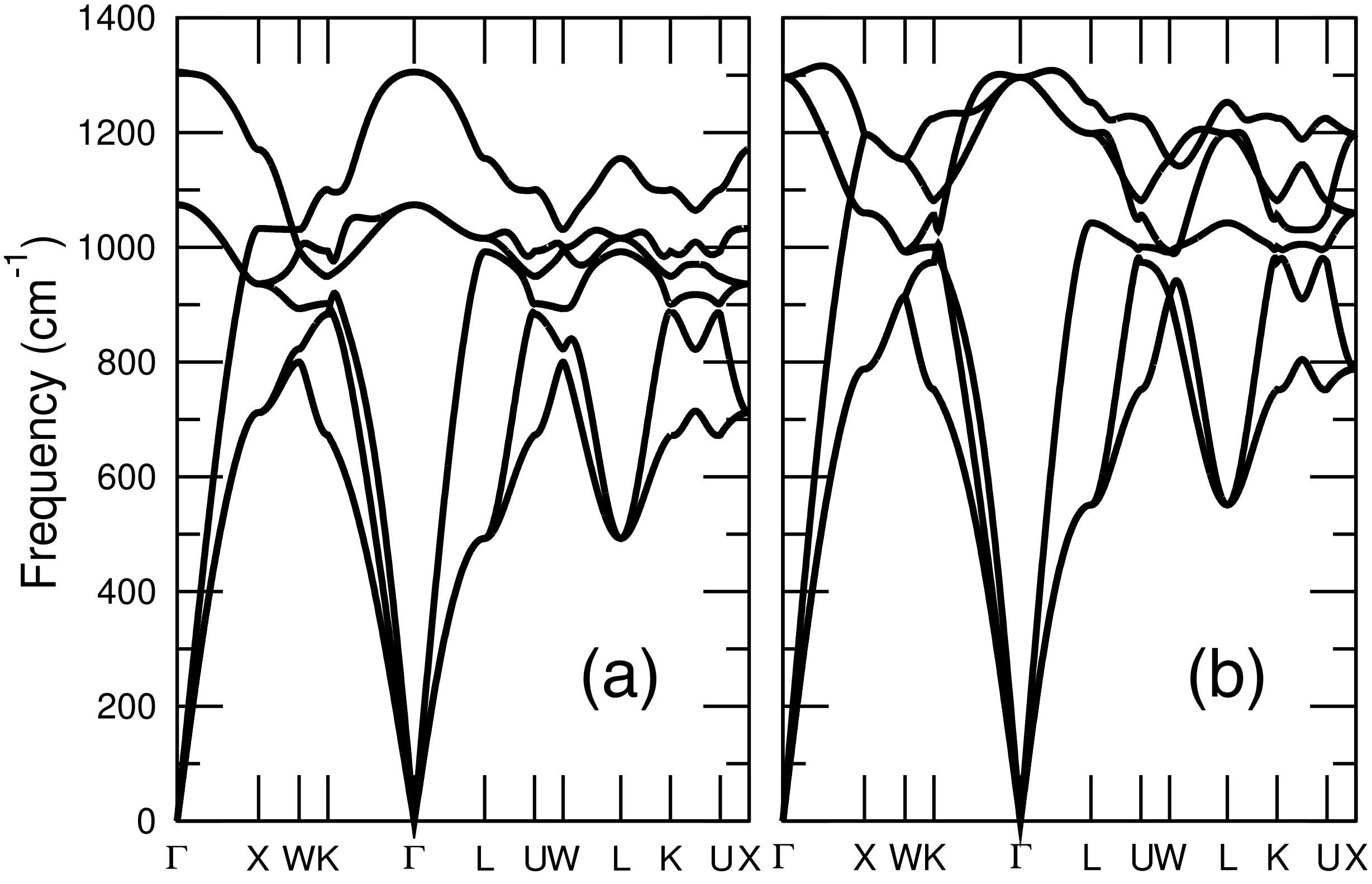
1400
1200
1000
800
600
400
200
0
Frequency (cm$^{-1}$)
(a)
(b)
Γ X WK Γ L UW L K UX
Γ X WK Γ L UW L K UX

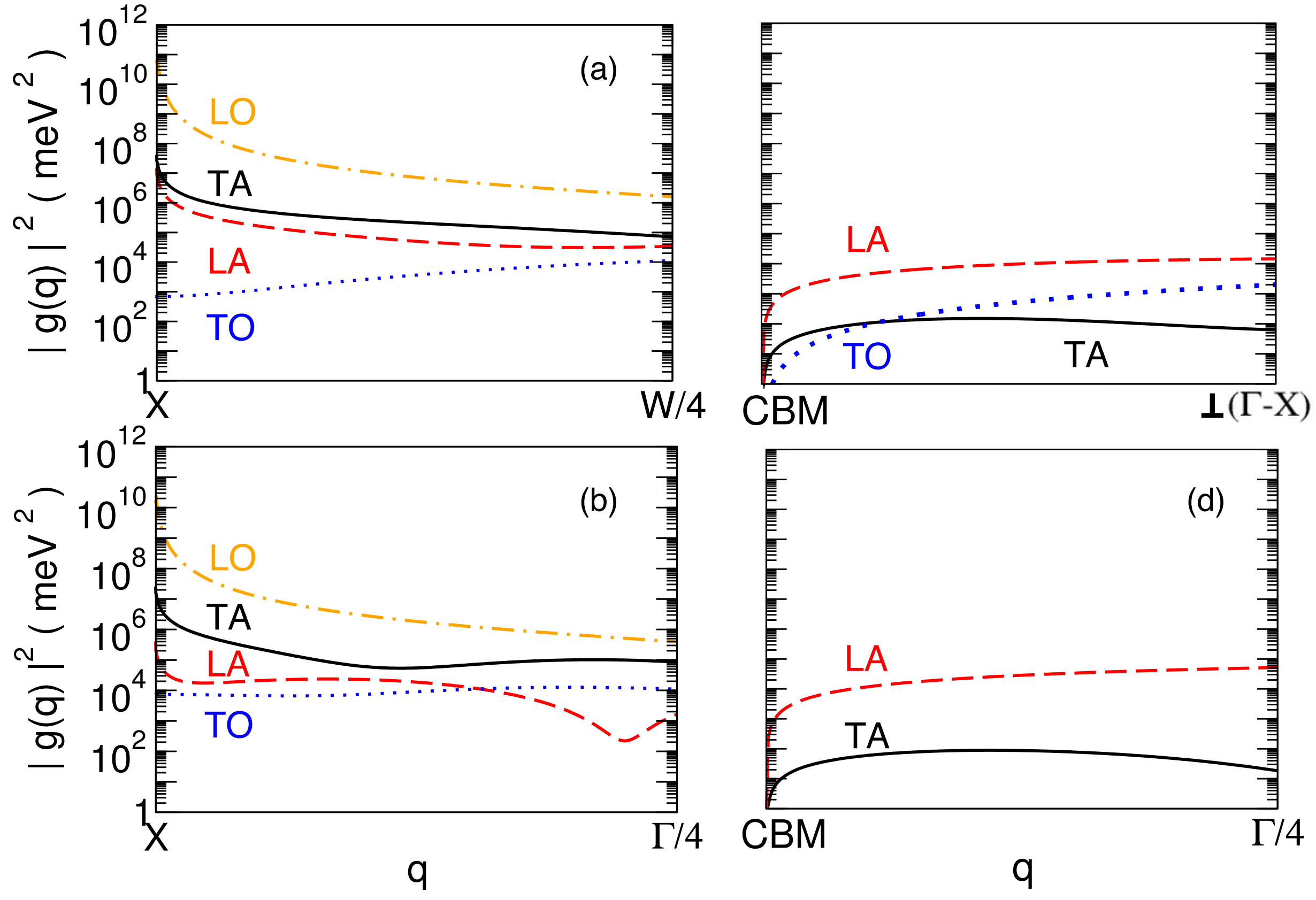

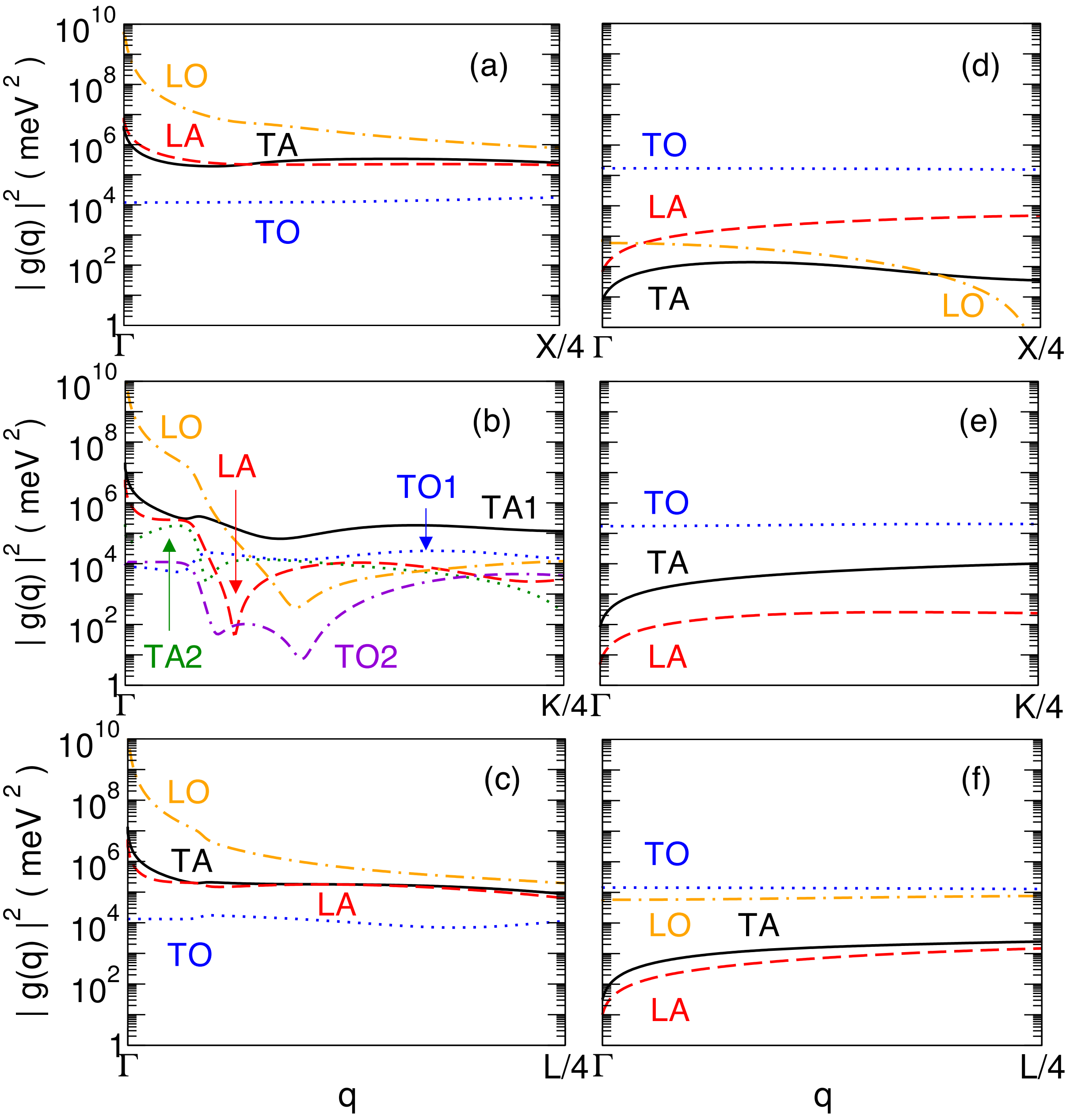

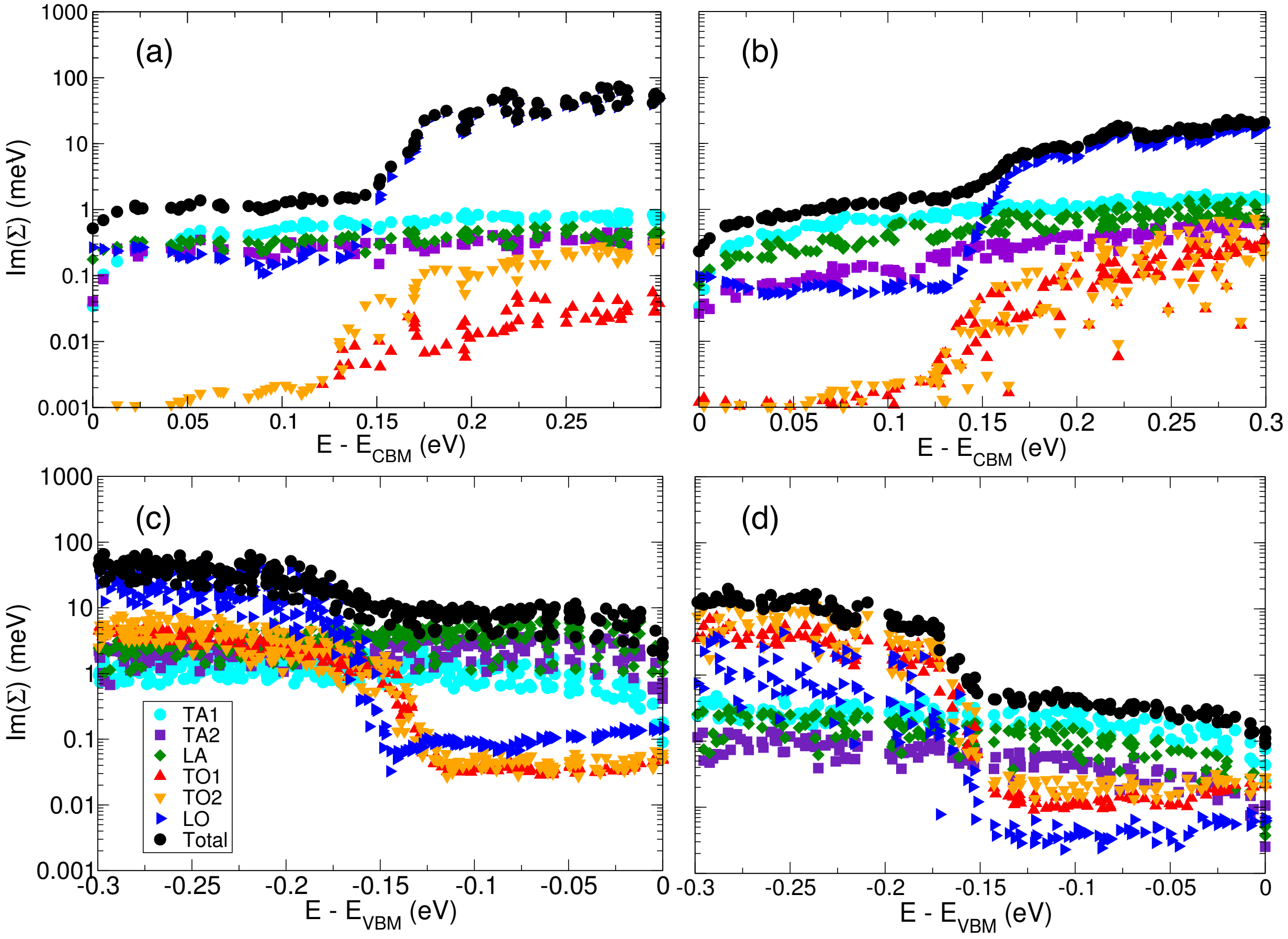

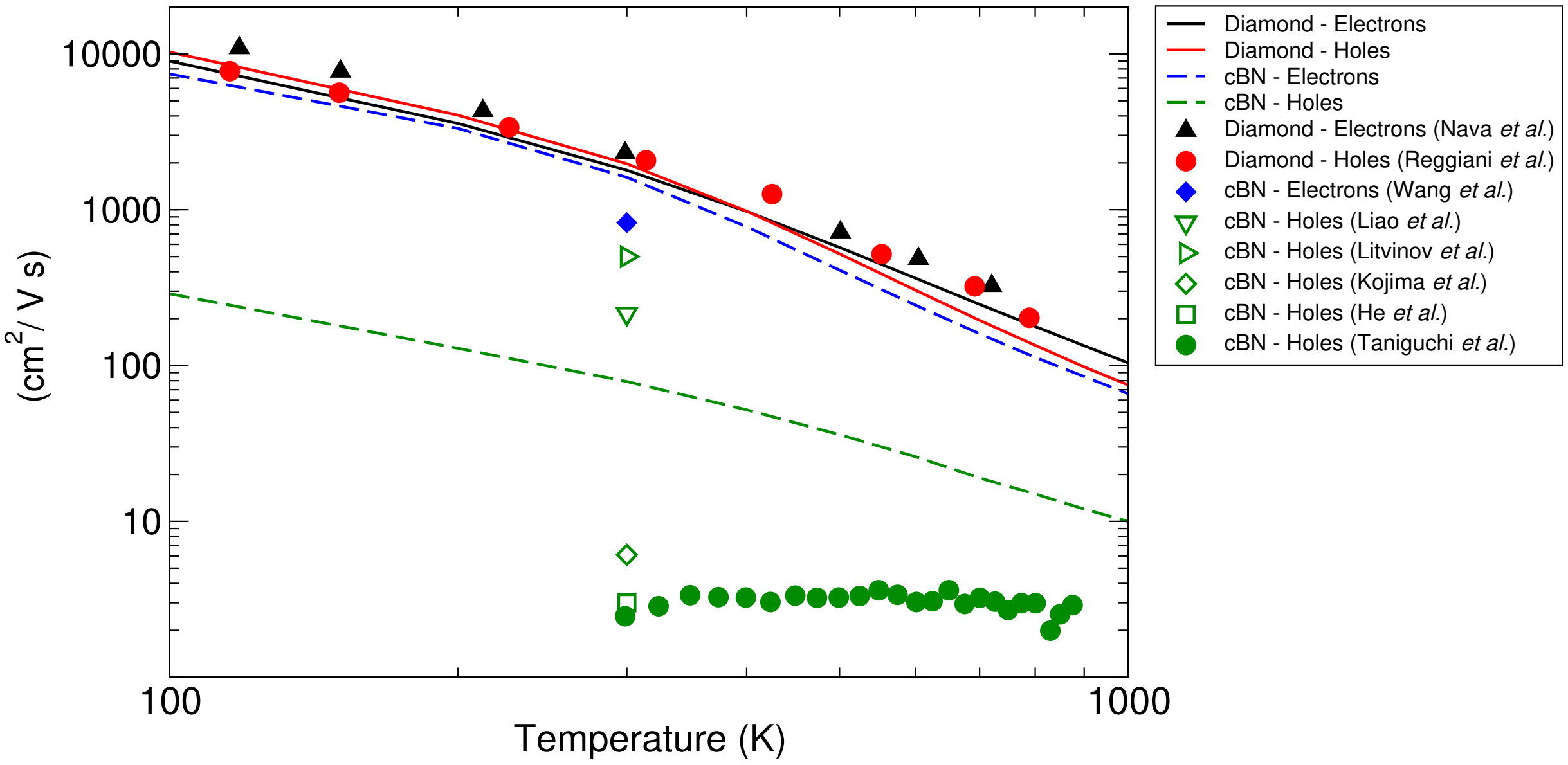
10000
1000
100
10
(cm$^2$/ V s)
100
1000
Temperature (K)
Diamond - Electrons
Diamond - Holes
cBN - Electrons
cBN - Holes
Diamond - Electrons (Nava *et al.*)
Diamond - Holes (Reggiani *et al.*)
cBN - Electrons (Wang *et al.*)
cBN - Holes (Liao *et al.*)
cBN - Holes (Litvinov *et al.*)
cBN - Holes (Kojima *et al.*)
cBN - Holes (He *et al.*)
cBN - Holes (Taniguchi *et al.*)

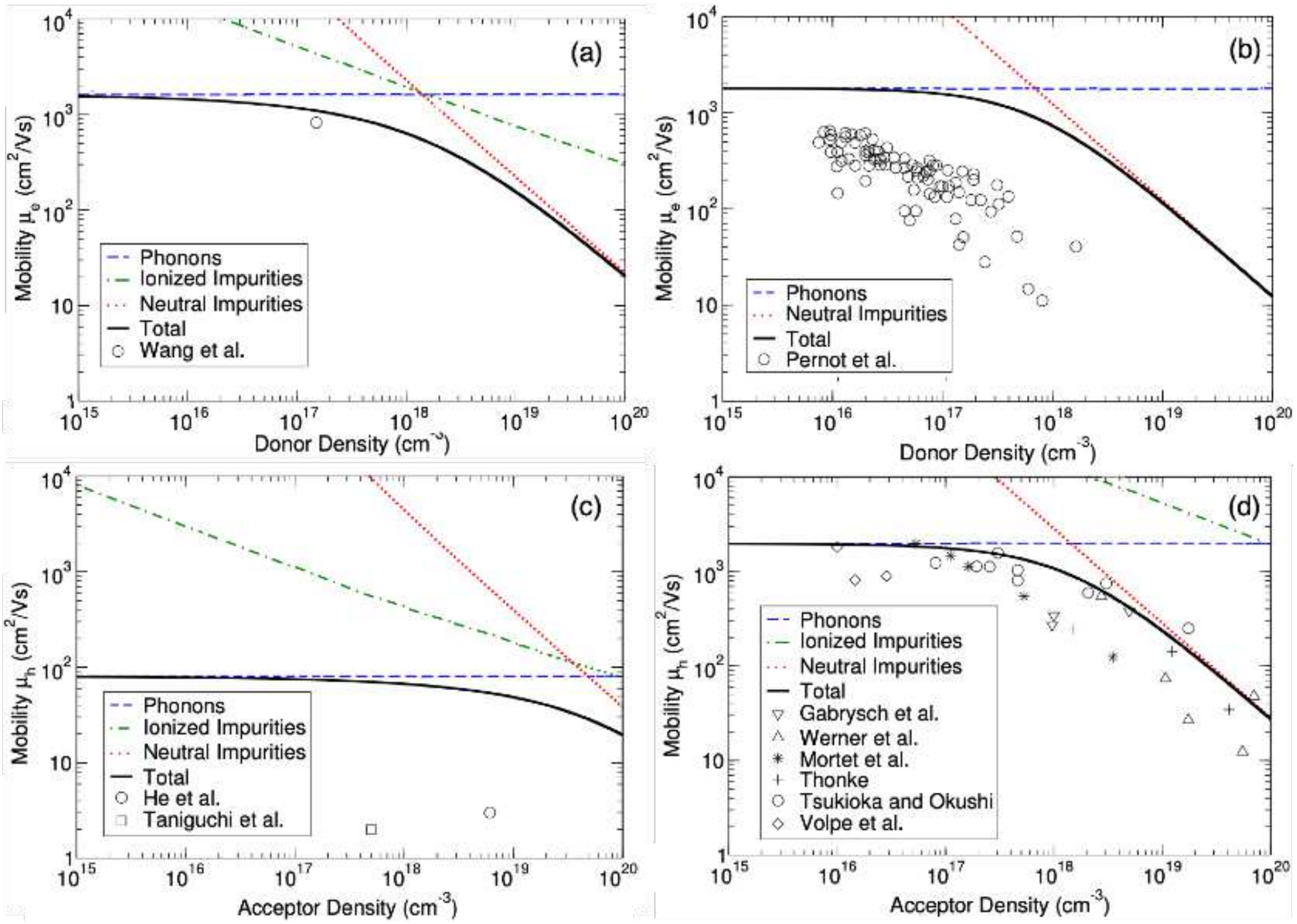